\documentclass[
reprint,
longbibliography,
amsmath, 
amssymb, 
aps, 
superscriptaddress,
prx,
]
{revtex4-2}
\usepackage{graphicx}% Include files
\usepackage{dcolumn}% Align table columns on decimal point
\usepackage{bm}% bold math
\usepackage{braket} %For the representation of bra and ket
\usepackage{hyperref}% add hypertext capabilities
\usepackage{subfigure} % subgraficos
\usepackage{nccmath}
\usepackage{amsthm} % Theorem Formatting
\usepackage{mathrsfs} %for the Hilbert symbol
\usepackage{xfrac} %for the fraction 1/2
\usepackage{pifont}
\usepackage[export]{adjustbox}
\usepackage{bbm} %probability font
\usepackage[utf8]{inputenc} 
\usepackage[normalem]{ulem}
\usepackage[dvipsnames,svgnames,table]{xcolor}
\usepackage[sort&compress]{natbib}
\usepackage{color}
\usepackage{longtable}
 
\usepackage{commath}
\usepackage{multirow}
\usepackage{array}

\newcommand{\ar}{\textcolor[rgb]{0,0,0}}
\begin{document}

\title{Benchmarking neutral atom-based quantum processors at scale}

 \author{Andrea B. Rava}
    \altaffiliation{a.rava@fz-juelich.de}
	\affiliation{Jülich Supercomputing Centre, Forschungszentrum Jülich, D-52425 Jülich, Germany}
    \affiliation{RWTH Aachen University, 52056 Aachen, Germany}

    \author{Kristel Michielsen}
	\affiliation{Jülich Supercomputing Centre, Forschungszentrum Jülich, D-52425 Jülich, Germany}
	\affiliation{Department of Computer Science, University of Cologne, 50931 Cologne, Germany }

  \author{J. A. Monta\~nez-Barrera}
    \altaffiliation{j.montanez-barrera@fz-juelich.de}
	\affiliation{Jülich Supercomputing Centre, Forschungszentrum Jülich, D-52425 Jülich, Germany}

\begin{abstract}
In recent years, neutral atom-based quantum computation has been established as a competing alternative for the realization of fault-tolerant quantum computation. However, as with other quantum technologies, various sources of noise limit their performance. With processors continuing to scale up, new techniques are needed to characterize and compare them in order to track their progress. In this work, we present two systematic benchmarks that evaluate these quantum processors at scale. We use the quantum adiabatic algorithm (QAA) and the quantum approximate optimization algorithm (QAOA) to solve maximal independent set (MIS) instances of random unit-disk graphs. These benchmarks are scalable, relying not on prior knowledge of the system's evolution but on the quality of the MIS solutions obtained. \ar{Rather than isolating individual sources of noise, they provide an application-level figure of merit: improvements in a QPU and its implementation of the protocols should be reflected in higher-quality MIS solutions.} We benchmark \texttt{quera\_aquila} and \texttt{pasqal\_fresnel} on problem sizes up to 102 and 85 qubits, respectively. Overall, \texttt{quera\_aquila} performs better on QAOA and QAA instances. Finally, we generate MIS instances of up to 1000 qubits, providing scalable benchmarks for evaluating future, larger processors as they become available. \ar{The proposed protocols can therefore serve as a common reference for comparisons, allowing future work to assess whether advances in hardware translate into measurable improvements in MIS solution quality.}

\begin{description}
	\vspace{0.2cm}
	\item[Keywords] Quantum Benchmarking, Aquila, Fresnel, neutral-atoms, Quantum Adiabatic Algorithm.
\end{description}
\end{abstract}

\maketitle
\section{Introduction}\label{Sec:Introduction}

Neutral atoms have rapidly emerged as one of the leading platforms for quantum computation and simulation, offering a unique combination of scalability, reconfigurability, and interaction control \cite{browaeys2020many,bernien2017probing,bluvstein2021controlling,cheng2024emergent,xu2023constantoverhead}. In these systems, individual atoms are trapped in optical tweezers or lattices and manipulated using laser light, while interactions are mediated through Rydberg excitations or spin-exchange mechanisms. This platform supports both analog and digital (gate-based) paradigms of quantum computation. In the analog mode, the dynamics of the system directly emulate the time evolution of a target Hamiltonian, making neutral atoms particularly suited for solving combinatorial optimization problems, quantum many-body dynamics, and quantum phase transitions \cite{Saffman2010_QINA,Browaeys_2020MBPNA,Ebadi_2022,Bombieri_2025, schuetz2025qredumisquantuminformedreductionalgorithm}. In parallel, advances in coherent control, qubit initialization, and high-fidelity Rydberg-mediated entangling gates have also enabled gate-based neutral-atom quantum computing \cite{Wagner_2024BenchNA} with the demonstration of fault-tolerant elements \cite{Sales_Rodriguez_2025, Zhou_2025}. 

Several commercial and academic efforts, including QuEra Computing~\cite{QuEraComputing}, Pasqal~\cite{Pasqal}, Atom Computing~\cite{AtomComputingTech}, and Infleqtion~\cite{Infleqtion}, are deploying large-scale neutral-atom processors with hundreds of atoms and near-term prospects for scaling to thousands~\cite{Henriet_2020Pasqal,wurtz2023AquilaQuEra}. In their analog mode, these programmable Rydberg simulators enable flexible encodings of the MIS via effective Ising Hamiltonians. As device sizes grow and performance claims increasingly rely on analog optimization, benchmarking becomes essential, motivating complementary work on improved annealing schedule design~\cite{Finzgar2024Bayesian}, strong classical baselines that reassess reported quantum speedups for MIS on unit-disk graphs~\cite{Andrist2023MIS}, and instance-reduction techniques that distinguish intrinsically hard from easy Rydberg-native MIS problems~\cite{Schuetz2025Compilation}.

As quantum processors continue to grow in complexity and scale, the demand for systematic benchmarking protocols has become increasingly important. Benchmarking in quantum computing generally serves two complementary purposes: (i) to quantify the performance and noise characteristics of a given hardware platform, and (ii) to provide a fair, architecture-independent means of comparison between devices. Traditional approaches, such as randomized benchmarking (RB) \cite{Knill_2008}, cross-entropy benchmarking (XEB) \cite{Boixo_2018}, and gate-set tomography \cite{Nielsen_2021}, focus on gate fidelities and control errors, providing microscopic performance measures for gate-based architectures \cite{blumekohout2013Tomography, Magesan2011BenchQP}. However, for analog neutral-atom processors, benchmarking must go beyond individual gate errors and instead evaluate the algorithmic quality of analog evolution under realistic many-body Hamiltonians.

Recent works have started to address this challenge by using problem-based or dynamical benchmarks, for instance, measuring the success probability of adiabatic protocols or the fidelity of analog quantum simulations of known spin models \cite{Ebadi_2021QPM_QUERA, Bluvstein_2022QP_Neut}. However, there is a need  for a unified framework for comparing the analog performance of large-scale devices.

In this work, we introduce a scalable benchmarking methodology that evaluates analog neutral-atom quantum processors through the performance of the QAA \cite{Keating_2013} and the QAOA \cite{tibaldi2025analogqaoabayesianoptimisation} to solve MIS instances of random unit-disk graphs. Our approach measures how effectively the analog dynamics of the system produce high-quality solutions to well-defined computational tasks. We apply this benchmark to QuEra’s Aquila \cite{wurtz2023aquila} and Pasqal’s Fresnel \cite{Pasqal2022HPCQS}, solving MIS problems up to 102 qubits and 85 qubits, respectively. We further generate benchmark instances for up to 1,000 qubits, providing a scalable testbed for assessing future devices as they grow in size and complexity.

\ar{The proposed protocol should not be interpreted as a complete characterization protocol for analog quantum hardware analogous to randomized benchmarking for gate-based devices. Rather, it is intended as an application-oriented benchmark that quantifies end-to-end performance for native Rydberg optimization problems under realistic operating conditions.}

\begin{figure*}
    \centering
    \includegraphics[width=1\linewidth]{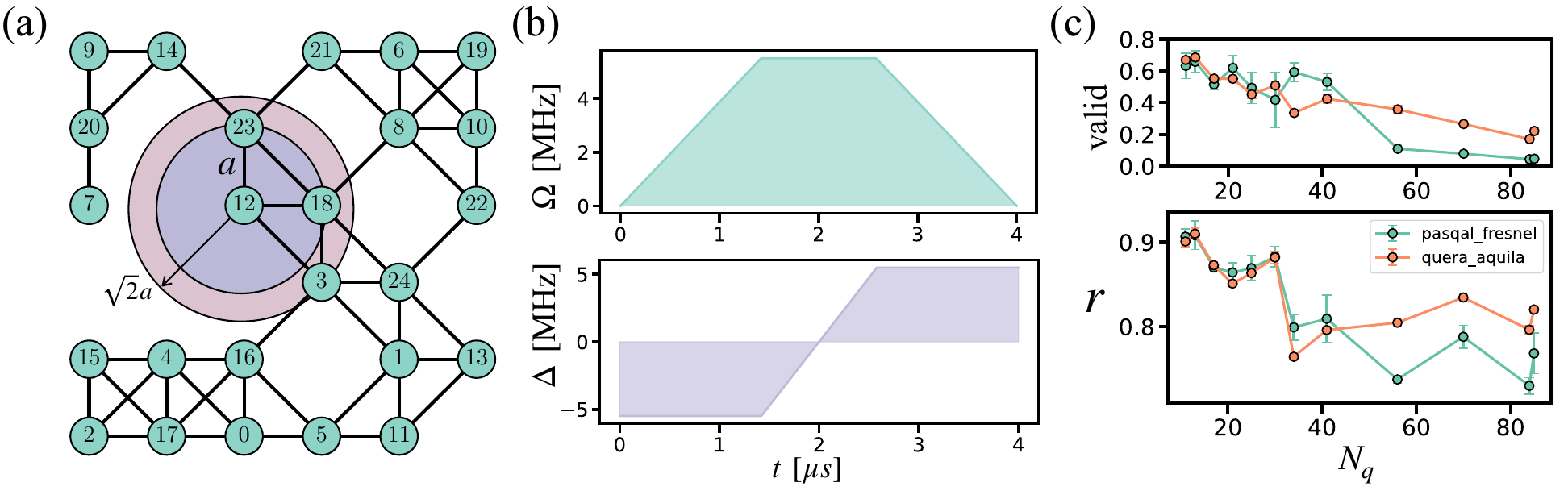}
    \caption{Benchmarking of neutral atom quantum devices. (a) Arrangement of the neutral atoms for the solution of a 25-qubit instance. $a$ is the minimum distance between atoms, and $\sqrt{2}a$ is the distance where a Rydberg interaction is considered strongly enough to encode an edge of the MIS. (b) QAA protocol for a $t=4\,\mathrm{\mu s}$. (c) Valid solutions (up) and approximation ratio (down), $r$, for problem sizes between 11 and 85 qubits for \texttt{quera\_aquila} and \texttt{pasqal\_fresnel}. Error bars represent the standard deviation for 3 experiments of 500 samples each.}
    \label{fig:IntroFigPaper}
\end{figure*}

Figure~\ref{fig:IntroFigPaper} summarizes this approach. In Fig.~\ref{fig:IntroFigPaper}(a), the arrangement of the atoms and the considered MIS problem is encoded. (b) The QAA schedule, if implemented correctly and depending on the noise level, should yield high-quality solutions to the MIS problem. (c) Experimental results on the two quantum processing units (QPUs) tested in terms of the proportion of valid solutions found, the approximation ratio, $r$, versus the number of qubits $N_q$.

\ar{The protocol evaluates the combined performance of hardware and algorithm rather than attempting to isolate hardware imperfections from algorithmic limitations.}

The paper is organized as follows. Section~\ref{Sec:Methods} provides a description of MIS, QAA, QAOA, and the experimental setup. In Sec.~\ref{Sec:Results}, the results of QAA and QAOA on real QPUs are presented. Finally, Sec.~\ref{Sec:Conclusions} provides some conclusions.

\section{Methods}\label{Sec:Methods}

\subsection{Maximal independent set problem}

In graph theory, an independent set is a collection of vertices in a graph such that no two are connected by an edge. The task of finding the Maximum Independent Set (MIS), the largest such subset, is a well-known NP-hard problem \cite{hespe2021targetedbranchingmaximumindependent}, meaning that no efficient (i.e., polynomial-time) algorithm is known to solve it for arbitrary graphs. For the MIS problem on a graph with $N_q$ vertices and edge set $E$, the cost function is
\begin{equation}\label{eq:cost_fun}
C(\vec{x}) = \sum_{i = 1}^{N_q} x_i - \lambda_0 \sum_{(i,j) \in E} x_i x_j,
\end{equation}

where $\lambda_0$ is a penalty coefficient to avoid solutions that do not fulfill the MIS constraints.

 The quantity used to evaluate a solver ability to find good solutions to this problem is known as the approximation ratio, $r$. It is defined as 
\begin{equation} 
\label{eq:approximation_ratio} 
r = \frac{\sum_i^n C(\vec{x_i})-C_{\text{worst}}}{C_{\text{opt}}-C_{\text{worst}}}, 
\end{equation}
where $\vec{x_i}$ are the $n$ solutions obtained with the solver, $C_{\text{opt}}$ and $C_{\text{worst}}$ are the optimal and worst possible values of $C(\vec x)$ defined in Eq.~\ref{eq:cost_fun}. This ratio lies in the interval $0 \leq r \leq 1$, where higher values indicate better solutions on average. In our case, we only consider the valid solutions found and therefore, $C_\mathrm{worst} =0$.

Another useful performance metric is the success probability, probability($x^*$), defined as the fraction of measurement outcomes that correspond to the optimal solutions $\{x_{i^*}\}$.

The quantum version of the objective function is defined in a $2^N_q$-dimensional Hilbert space with computational basis states $\ket{z} \in \{\ket 0,\ket 1\}^{\otimes N_q}$. The classical cost function is converted to a diagonal Hamiltonian $\hat{H}^C$ by mapping each bit $x_i$ to the occupation operator $\hat n = (1 + \hat\sigma_i^z)/2$, 

\begin{equation} \label{eq:cost_H} 
\hat H^C = \sum_i^{N_q}\left(w_i\hat n_i+\sum_{j < i} l_{ij} \hat n_i\hat n_j\right), 
\end{equation}
where $w_i$ and $l_{ij}$ are real-valued weights.
 
The MIS constraints map naturally onto the Rydberg blockade phenomenon in neutral-atom quantum computing. Specifically, the Rydberg blockade prevents two nearby atoms from being simultaneously excited to a Rydberg state $\ket{r}$, mirroring the MIS condition that connected vertices cannot both be included in the independent set \cite{pichler2018quantumoptimizationmaximumindependent}. The Rydberg Hamiltonian that describes this phenomenon is given by

\begin{equation}\label{eq:HRyd}
\hat{H}_{\mathrm{Ryd}} = \sum_i^{N_q} \left[ \frac{\Omega(t)\hbar}{2} \hat{\sigma}_i^x - {\Delta(t)\hbar} \hat{n}_i + \sum_{j<i} V_{ij} \hat{n}_i \hat{n}_j \right],
\end{equation}
where $V_{ij} = C_6/R^6_{ij}$ is the interaction between atoms $i$ and $j$ at distance $R_{ij}$, and $C_6$ is the Rydberg interaction constant.

This correspondence allows us to encode the MIS problem directly into the ground state of the Rydberg Hamiltonian. For certain geometric graphs, particularly unit disk graphs (UDGs), where vertices represent points in the plane and edges connect pairs within a fixed distance, this encoding can be achieved without any overhead in the number of qubits, using each atom to represent a single graph vertex \cite{Ebadi_2022}.

In this work, we use Diagonal-connected Unit-disk Grid Graphs (DUGGs), such as the one in Fig.~\ref{fig:IntroFigPaper}(a). Despite their constrained and regular structure, finding the MIS for such graphs remains NP-hard \cite{Ebadi_2022} and a non-trivial task on neutral atom-based quantum processing units (QPUs).

We arrange the atoms on a square grid such that both nearest neighbors (horizontal and vertical) and next-nearest neighbors (diagonal) fall within the unit disk interaction range. As a result, in the associated graph, edges connect both adjacent and diagonally neighboring vertices. To generate a variety of graph instances, we apply a random dropout process to square grids of varying sizes, controlling the density by adjusting the probability that each site is occupied by an atom.

\subsection{QAA algorithm}\label{sec:QAA_alg}

The core idea of the QAA (also known as Quantum Annealing) is to exploit the adiabatic theorem: if a quantum system starts in the ground state of an initial Hamiltonian and evolves sufficiently slowly, it will remain in the instantaneous ground state of the time-dependent Hamiltonian throughout the evolution~\cite{zunkovič2024VarQAA}. By designing the final Hamiltonian so that its ground state encodes the solution to a desired optimization problem, the system ideally ends up in that solution state at the end of the evolution.

For Rydberg-atom-based quantum devices, this adiabatic evolution can be naturally implemented using the tunable Rydberg Hamiltonian parameters. Since the initial state of the device is prepared with all atoms in their electronic ground state $\ket{0}$, we choose the initial Hamiltonian such that this state is indeed its ground state. This is readily achieved by setting the initial detuning to a large negative value, $\Delta(t=0) = \Delta_\mathrm{in} < 0$, and keeping the Rabi driving field off, i.e., $\Omega(t=0) = 0$. In this configuration, the ground state corresponds trivially to all qubits in $\ket{0}$.

The adiabatic evolution proceeds in three main stages (See Fig.~\ref{fig:IntroFigPaper}(b)):

\begin{itemize}
    \item \textbf{Ramp-up of the driving field}: Over an initial time interval $t_\mathrm{rise}$, the Rabi frequency $\Omega(t)$ is increased linearly from 0 to a chosen maximum value $\Omega_\mathrm{max}$, while keeping the detuning $\Delta$ fixed at $\Delta_\mathrm{in}$. This activates coherent transitions that enable superpositions and tunneling between different classical configurations.
    \item \textbf{Sweeping the detuning}: During a period $t_\mathrm{sweep}$, the driving field is held constant at $\Omega_\mathrm{max}$, while the detuning $\Delta(t)$ is linearly swept from the initial negative value $\Delta_\mathrm{in}$ to a final positive value $\Delta_\mathrm{fin}$. This sweep biases the Hamiltonian towards the configurations that solve the MIS.
    
    \item \textbf{Ramp-down of the driving field}: In the final interval $t_\mathrm{fall}$, the detuning $\Delta(t)$ is kept fixed at $\Delta_\mathrm{fin}$, while the driving field $\Omega(t)$ is linearly ramped back down to zero. This effectively suppresses further quantum fluctuations.
\end{itemize}

\subsection{QAOA algorithm}\label{QAOA_alg}

In general, an optimization problem can be formulated as minimizing a cost function $C(\vec x)$ that encodes the problem’s constraints and depends on a binary variable vector $\vec x = (x_0, x_1, \ldots, x_{N_q})^{\sf T}$, with $x_i \in \{0,1\}$. The goal is to find the bitstring $\vec x^*$ that minimizes $C(\vec x)$.

QAOA prepares a quantum state with high probability of yielding a near-optimal bitstring $\vec{x}$ by optimizing over variational parameters. The procedure consists of:

\begin{enumerate}
    \item Initial state: Prepare the uniform superposition $\ket{s} = \ket{+}^{\otimes N_q}.$
    \item Variational evolution: Apply $p$ layers of alternating unitaries generated by $\hat{H}^C$ and a mixing Hamiltonian $\hat{H}^B$, parameterized by $(\vec{\gamma}, \vec{\beta})$    
\begin{equation} \label{eq:final_state}
\ket{\vec{\gamma}, \vec{\beta}} = \hat{U}_p^M \hat{U}_p^C \cdots \hat{U}_1^B \hat{U}_1^C \ket{s},
\end{equation}
where $\hat{U}_k^C = e^{-i\gamma_k \hat{H}^C}$ and $\hat{U}_k^M = e^{-i\beta_k \hat{H}^M}$ and the mixing Hamiltonian is $\hat{H}^M = \sum_{j=1}^{N_q} \hat\sigma^x_j$.
    \item Measurement: Estimate the expectation value of the cost Hamiltonian
\begin{equation} \label{eq:F_p}
F_p(\vec{\gamma}, \vec{\beta}) = \bra{\vec{\gamma}, \vec{\beta}} \hat{H}^C \ket{\vec{\gamma}, \vec{\beta}},
\end{equation}    
using $n_\mathrm{shots}$ samples:
\begin{equation} \label{eq:statistic_F_p}
F_p = \frac{\sum_{i=1}^{n_\mathrm{shots}} w_i C(\vec x_i)}{n_\mathrm{shots}},
\end{equation}
where $w_i$ counts occurrences of bitstring $\vec x_i$.
\item Classical optimization: perform steps $1-3$ in an optimization loop to minimize $F_p$ over $(\vec{\gamma}, \vec{\beta})$. In this work, we used the Nelder-Mead simplex method \cite{Nelder} as classical optimizer.
\end{enumerate}

The correspondence between QAOA cost Hamiltonian, Eq.\ref{eq:cost_H}, $\hat{H}^M$, and the Rydberg Hamiltonian, Eq. \ref{eq:HRyd} can be described as follows. The $\hat{n}_i$ and $\sum_{j <i}V_{ij}\hat n_i \hat n_j$ correspond to the cost function, while $\hat{\sigma}_i^x$ acts as a mixing term. The QAOA parameters $\gamma_k$ and $\beta_k$ correspond to different values $\Delta_k$ and $\Omega_k$, respectively. 

However, implementing QAOA using neutral atoms differs in three key ways. First, the initial state is the all-down state $\ket{0}^{\otimes N_q}$. Second, the interaction term is always active, making it impossible to isolate the mixing Hamiltonian during evolution. Finally, in our implementation, the $\Omega(t)$ and $\Delta(t)$ pulses act simultaneously, rather than in the alternating layers characteristic of QAOA.

\subsection{Transfer learning on QAOA}

Transfer learning (TL) refers to reusing pre-optimized QAOA parameters from one problem instance on different instances \cite{Monta_ez_Barrera_2025}, thereby reducing the need for costly classical optimization for each new problem. This means applying pre-trained parameters $\vec\gamma = (\gamma_1, \gamma_2, \ldots, \gamma_{p})$ and $\vec\beta = (\beta_1, \beta_2, \ldots, \beta_{p})$ to problem instances not used during the original optimization. The approach involves first optimizing the parameters for a specific instance, then evaluating their performance on other instances.

We optimized QAOA parameters for $p=10$ and total evolution time $t_{\mathrm{tot}} = 2 \, \mathrm{\mu s}$ simultaneously on 12 randomly generated DUGGs with system sizes between 10 and 12 qubits, using the JUelich Quantum Annealing Simulator (JUQAS), which solves the Schrödinger equation using the Suzuki-Trotter product-formula algorithm \cite{suzuki1976,suzuki84,trotter59,deraedt87,huyghebaert1990}.

% The time evolution of the system was emulated by solving the time-dependent Schrödinger equation (with $\hbar =1$)

% \begin{equation}
% i\frac{\partial}{\partial t}|\psi(t)\rangle =H(t)|\psi(t)\rangle
% \end{equation}

% numerically. Here, $|\psi(t)\rangle$ represents the state vector, while the Hamiltonian governing the system dynamics is given by

% \begin{equation}
% \begin{split}
% {H}(t) = \sum\limits_{i} \Big[\big( f_i^x(t)h_i^x\sigma_i^x + f_i^y(t)h_i^y\sigma_i^y + f_i^z(t)h_i^z\sigma_i^z \big) + \\
% +\sum\limits_{j<i} \big( F_{ij}^x(t)J_{ij}^x \sigma_i^x\sigma_j^x + F_{ij}^y(t)J_{ij}^y \sigma_i^y\sigma_j^y + F_{ij}^z(t)J_{ij}^z \sigma_i^z\sigma_j^z \big)\Big].    
% \end{split}
% \end{equation}

% The numerical approach employed for solving the Schrödinger equation is the Suzuki-Trotter product-formula algorithm \cite{suzuki1976,suzuki84,trotter59,deraedt87,huyghebaert1990}, which allows for full state-vector emulation.

\begin{figure}[!tbh]
\centering
\includegraphics[width=8cm]{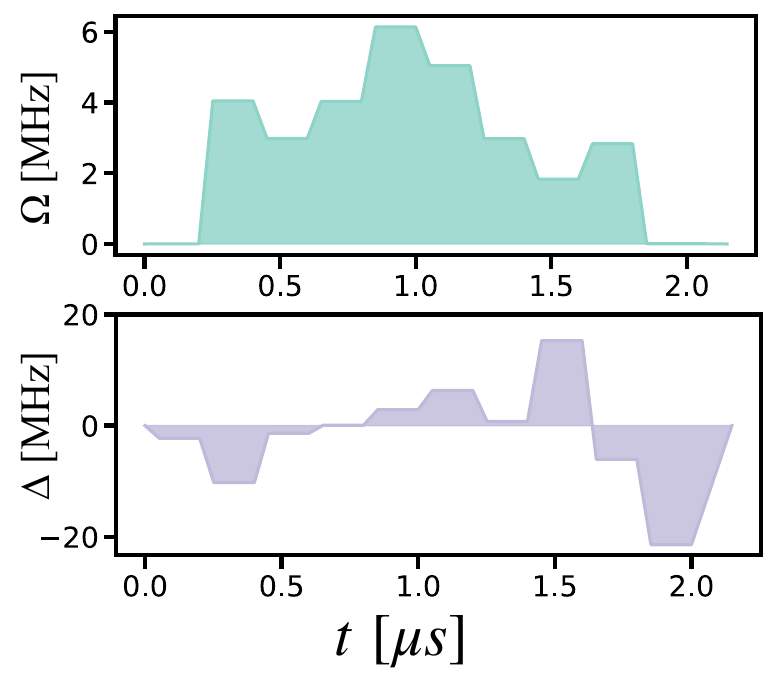}
\caption{\label{Fig:QAOA-protocol} Optimized schedule for the QAOA protocol.}
\end{figure}

For the emulations, the starting point was the annealing schedule of Fig.~\ref{fig:IntroFigPaper}(b), and the average over the 12 problems is taken to ensure that the parameters obtained work for different problems. The optimized parameter schedules is shown in Fig.~\ref{Fig:QAOA-protocol}.
% This involved numerically solving the time-dependent Schrödinger equation (with $\hbar = 1$),

% \begin{equation}
% i\frac{\partial}{\partial t}|\psi(t)\rangle = {\hat H}(t)|\psi(t)\rangle
% \end{equation}\label{eq:HamJuqas}
% where $|\psi(t)\rangle$ is the time-evolved quantum state and $\hat{H}(t)$ is given by,
% \begin{equation}
% \begin{aligned}
% \hat H(t) = &\sum\limits_{i=0}^{N-1} (f_i^x(t)h_i^x\sigma_i^x + f_i^y(t)h_i^y\sigma_i^y + f_i^z(t)h_i^z\sigma_i^z) \\
% &+ \sum\limits_{i<j} 
% \begin{aligned}(&F_{ij}^x(t)J_{ij}^x \sigma_i^x\sigma_j^x\\
% &+ F_{ij}^y(t)J_{ij}^y \sigma_i^y\sigma_j^y\\ &+ F_{ij}^z(t)J_{ij}^z \sigma_i^z\sigma_j^z).
% \end{aligned}
% \end{aligned}
% \end{equation}
% This general Hamiltonian supports the emulation of both quantum annealers and quantum simulators. We use the Suzuki-Trotter product-formula algorithm~\cite{suzuki1976,suzuki84,trotter59,deraedt87,huyghebaert1990} for full state-vector evolution. Our implementation supports decomposition into either single- and two-qubit terms or Pauli-basis terms ($\sigma^x$, $\sigma^y$, $\sigma^z$), and is GPU-accelerated via OpenACC. For larger system sizes ($\gtrapprox$ 30 qubits), distributed memory across multiple GPUs is required, with inter-GPU communication handled via CUDA-aware MPI, as in the Jülich Universal Quantum Computer Simulator (JUQCS)~\cite{deraedt07,deraedt18,Willsch2022_gpu}.

\subsection{Experimental setup}\label{sec:ExpSet}
% \colorbox{green}{---------------------}
% \\
% How parameters can be translated to a given QPU. Characteristics of each QPU are listed in a table with the most important aspects of the QPUs. Number of shots we use...
% \\
% \colorbox{green}{---------------------}

The experiments are executed on two neutral atom QPUs, Fresnel from Pasqal, \texttt{pasqal\_fresnel}, and Aquila from QuEra, \texttt{quera\_aquila} \cite{wurtz2023aquilaqueras256qubitneutralatom}. Their main specifications are listed in Table~\ref{tab:DevsSpecs}. For the experimental results, we took 500 samples for each problem size on three different days on each device for both the QAOA and QAA protocols. 

For QAOA parameters optimization, we used an emulator with an interaction coefficient $C_6 = 865723\,\mathrm{MHz}$, matching that of Fresnel. To apply the same schedule to Aquila, which has a different $C_6$, we ensure consistent interaction strengths by using the condition 
\begin{equation}\label{eq:RelIntDist_Devs}
C_6^{\mathrm{Fresnel}}/a_\mathrm{Fresnel}^6 = C_6^{\mathrm{Aquila}}/a_\mathrm{Aquila}^6,
\end{equation}
where $a$ is the distance between adjacent atoms, and for \texttt{pasqal\_fresnel} is used $a_\mathrm{Fresnel} = 5\,\mathrm{\mu m}$ which corresponds to $a_\mathrm{Aquila} = 6.79\, \mathrm{\mu m}$ for \texttt{quera\_aquila}.

For the QAA, we employed the schedule structure defined in Section \ref{sec:QAA_alg}. We used fixed values for the driving field and detunings: $\Omega_\mathrm{max} = 5.5 ,\mathrm{MHz}$, $\Delta_\mathrm{in} = -5.5 ,\mathrm{MHz}$, and $\Delta_\mathrm{fin} = 5.5 ,\mathrm{MHz}$. The total evolution time $t_\mathrm{tot}$ was varied across different experiments to investigate adiabaticity and solution quality. The ramp-up ($t_\mathrm{rise}$) and ramp-down ($t_\mathrm{fall}$) durations for $\Omega(t)$ were scaled accordingly with $t_\mathrm{tot}$ to maintain a symmetric profile. The detuning $\Delta(t)$ was linearly swept during a central interval $t_\mathrm{sweep}$, as described in Section \ref{sec:QAA_alg}.

\setlength{\tabcolsep}{0.5em} % for the horizontal padding
\renewcommand{\arraystretch}{1.2}% for the vertical padding
\begin{table}
\begin{center}
\begin{tabular}{||c|c|c||}
\hline
& Fresnel & Aquila \\
\hline
\hline
$C_6$ & $865723\,\mathrm{MHz}$ & $5420441 \,\mathrm{MHz}$\\
\hline
max num. atoms & $100$ & $256$\\
\hline
$a_\mathrm{min}$ & $5 \,\mathrm{\mu m}$  & $4 \,\mathrm{\mu m}$  \\
\hline
$t_\mathrm{max}$ & $6 \,\mathrm{\mu s}$  & $4 \,\mathrm{\mu s}$\\
\hline
$\Omega_\mathrm{max}$ & $2\times 2\pi \,\mathrm{MHz}$  & $2.5\times 2\pi\,\mathrm{MHz}$\\
\hline
$|\Delta|_\mathrm{max}$ & $7.75\times 2\pi \,\mathrm{MHz}$  & $20\times 2\pi\,\mathrm{MHz}$\\
\hline
% ... & ... & ... \\
% \hline
\end{tabular}
\caption{Specifications of \texttt{quera\_aquila} and \texttt{pasqal\_fresnel}.}
\label{tab:DevsSpecs}
\end{center}
\end{table}

Measurement noise and atom loss are major challenges in analog quantum computing, and the two platforms address them differently. In \texttt{pasqal\_fresnel}, only atoms that remain trapped are detected, so both Rydberg excitations and lost atoms appear as missing, creating ambiguity that must be modeled through SPAM parameters: $\eta$ for preparation errors, $\epsilon$ for false positives, and $\epsilon'$ for false negatives. In contrast, \texttt{quera\_aquila} provides detailed shot-level diagnostics, reporting whether readout errors occurred in each run. This enables post-selection to discard faulty bitstrings.

\section{Results}\label{Sec:Results}

\begin{figure}[!tbh]
\centering
\includegraphics[width=8cm]{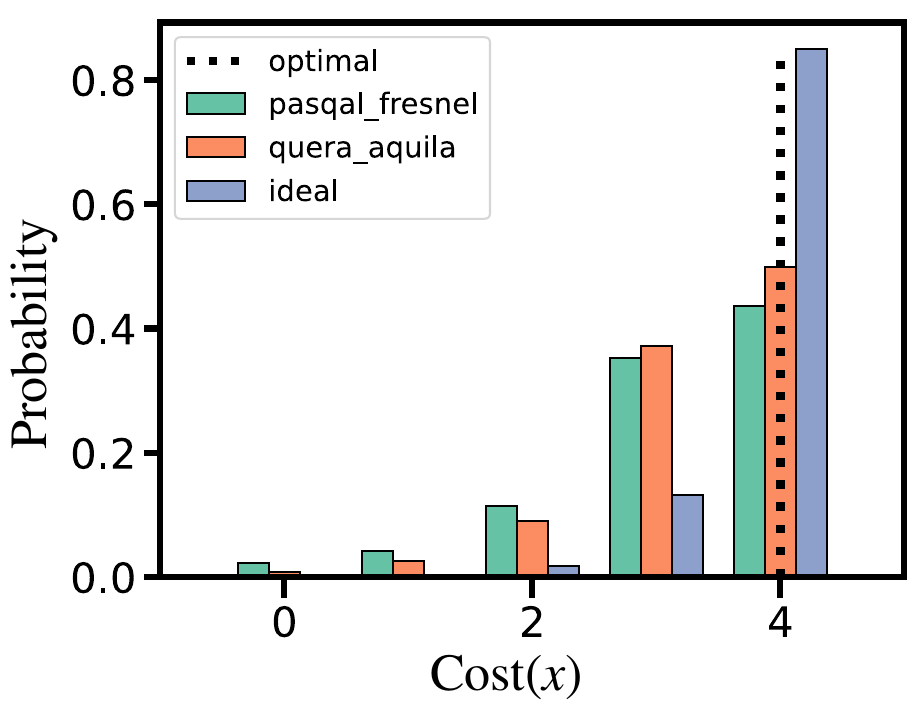}
\caption{\label{Fig:prob_13} Probability distribution of different solutions comparing \texttt{quera\_aquila}, \texttt{pasqal\_fresnel}, and an ideal (noiseless) simulator for a problem with 13 qubits using QAA with $t=4\,\mathrm{\mu s}$.}
\end{figure}

Figure~\ref{Fig:prob_13} shows the probability versus cost function of valid solutions for a 13-qubit MIS problem using \texttt{quera\_aquila}, \texttt{pasqal\_fresnel}, and an ideal simulator. The optimal solution in this case involves 4 vertices and both QPUs can find it with large probability over the 500 samples taken, but still is below the ideal result. This is a consequence of noise, and the majority of the solutions of both solvers are still valid, but they differ in one or two from the optimal.

Extended results of solution quality versus number of qubits are given in Table \ref{Tab:QAA} and Fig.~\ref{fig:IntroFigPaper}(c) which summarizes the results for system sizes up to $85$ qubits for the $4\,\mathrm{\mu s}$ QAA schedule, using both the approximation ratio and the fraction of valid solutions as performance metrics. For problem sizes $N_q \leq 30$, the two QPUs exhibit comparable behaviour. Although \texttt{pasqal\_fresnel} shows slightly better performance at $N_q = 34$ and $N_q = 41$, the overall trend indicates that \texttt{quera\_aquila} outperforms \texttt{pasqal\_fresnel}.

After a recent upgrade of \texttt{pasqal\_fresnel}, we were able to conduct experiments for $N_q > 41$. For these experiments, the quality of the solutions on \texttt{pasqal\_fresnel} drops to less than 11\% for  $N_q > 41$, which might indicate problems with the setup when operating the device at these scales.

\begin{table}[h!]
\centering

\begin{tabular}{|c|cc|cc|}
\hline
 & \multicolumn{2}{c|}{$r$} & \multicolumn{2}{c|}{valid}\\
\hline
$N_q$ & Fresnel & Aquila&
Fresnel & Aquila \\
\hline
11 & 0.9066 & 0.9008 & 0.632 & 0.669 \\
13 & 0.9082 & 0.9101 & 0.657 & 0.685 \\
17 & 0.8700 & 0.8726 & 0.515 & 0.551 \\
21 & 0.8641 & 0.8509 & 0.619 & 0.550 \\
25 & 0.8691 & 0.8634 & 0.493 & 0.452 \\
30 & 0.8828 & 0.8819 & 0.417 & 0.508 \\
34 & 0.7991 & 0.7641 & 0.593 & 0.335 \\
41 & 0.8092 & 0.7959 & 0.530 & 0.424 \\
56 & 0.7372 & 0.8045 & 0.109 & 0.357 \\
70 & 0.7877 & 0.8344 & 0.078 & 0.265 \\
84 & 0.7297 & 0.7962 & 0.043 & 0.170 \\
85 & 0.7682 & 0.8201 & 0.046 & 0.221 \\
\hline
\end{tabular}
\caption{QAA for $t=4\,\mathrm{\mu s}$ approximation ratio and valid solution fractions for \texttt{pasqal\_fresnel} and \texttt{quera\_aquila}.}
\label{Tab:QAA}
\end{table}

\begin{figure}[!tbh]
\centering
\includegraphics[width=8cm]{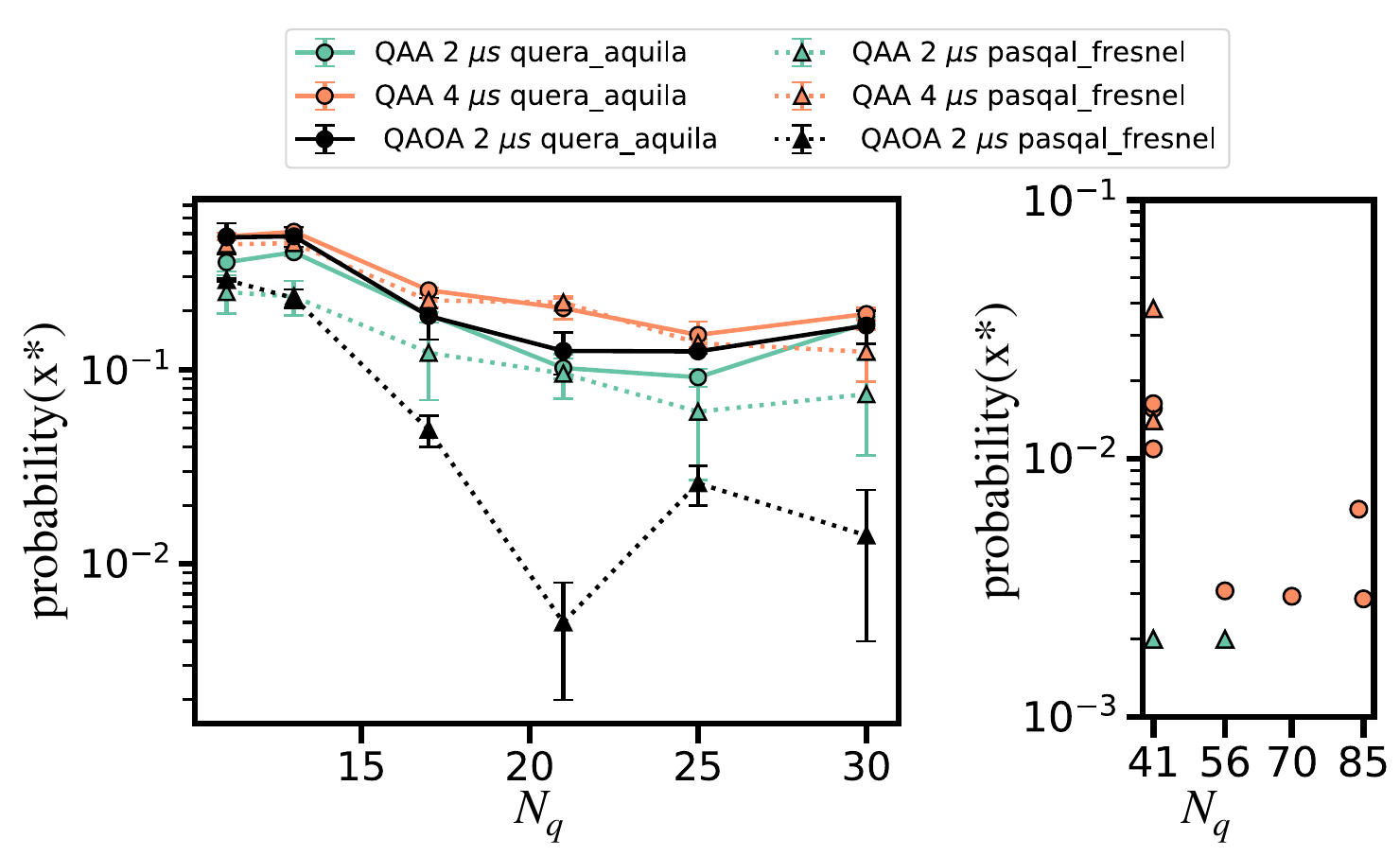}
\caption{\label{Fig:prob_summary} Probability of success vs. number of qubits for QAOA and QAA on \texttt{quera\_aquila} and \texttt{pasqal\_fresnel}. (left) $N_q\le 30$ (right) $N_q>30$.}
\end{figure}

Figure~\ref{Fig:prob_summary} compares QAOA and QAA performance for both QPUs. For the $2\,\mathrm{\mu s}$ schedule, \texttt{quera\_aquila} consistently outperforms \texttt{pasqal\_fresnel}. In particular, \texttt{pasqal\_fresnel} is less likely to find optimal solutions under QAOA protocol, an indication that implementing rapid variations of the driving fields might cause a degradation in the solution quality.

For QAOA, neither device finds the optimal solution for $N_q > 30$ qubits , which is a consequence of sampling size, difficulty in implementing the algorithm on the QPUs, and the own performance of QAOA at $2 \,\mathrm{\mu s}$. Fig.~\ref{Fig:prob_summary}(right) shows the QAA results at $N_q > 30$. For problem sizes up to 56 qubits,  \texttt{pasqal\_fresnel} finds the optimal solution even at $2\,\mathrm{\mu s}$ while \texttt{quera\_aquila} finds it at least once in three experiments for up to $N_q=85$.

\begin{figure}[!tbh]
\centering
\includegraphics[width=8cm]{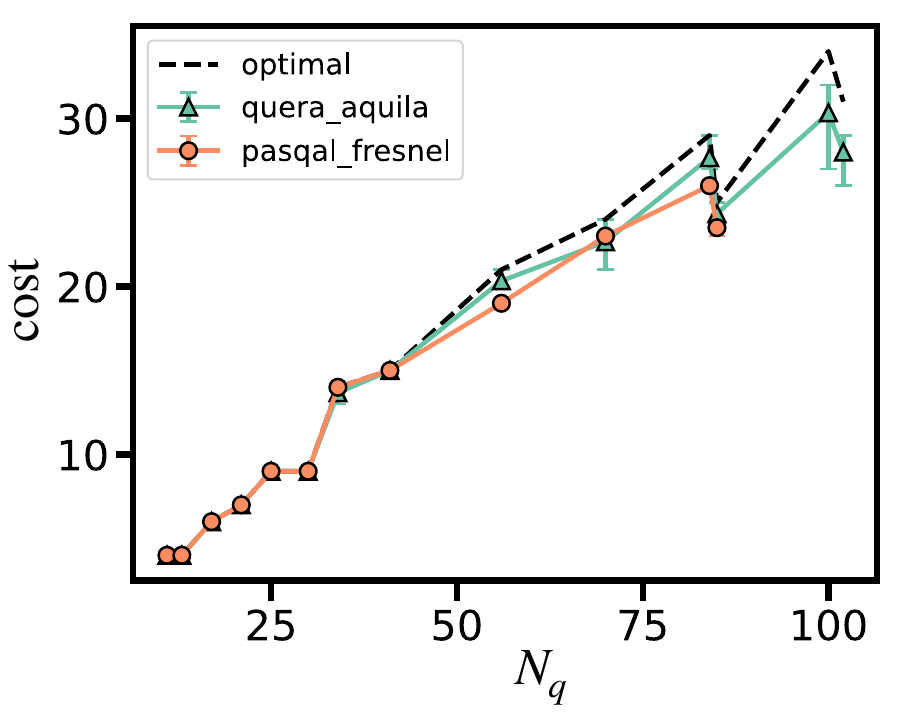}
\caption{\label{Fig:Cost_scaling} Cost of the best solution found versus number of qubits for QAA and $t=4\,\mathrm{\mu s}$. The dashed line represents the optimal cost for each instance. The error bars represent the standard deviation over the 3 experiments.}
\end{figure}

Finally, Figure~\ref{Fig:Cost_scaling} shows the  QAA best result obtained for each QPU at $t=4\,\mathrm{\mu s}$. We could not implement problems larger than $N_q > 85$ on \texttt{pasqal\_fresnel}, so there are no points above it. Up to $~56$ qubits, both QPUs find the optimal solution at least once. As the system size increases, the probability of finding the optimal solution  gradually degrades, which is expected because the problems are more difficult to solve, and noise might impact the solution quality.

% \begin{figure}[!tbh]
% \centering
% \includegraphics[width=8cm]{preparation_error.pdf}
% \caption{\label{Fig:preparation_error} Percentage of preparation errors in Aquila.}
% \end{figure}

\section{Discussion and Conclusions}\label{Sec:Conclusions}

In this work, we have presented two benchmarking strategies for neutral atom-based quantum computers that can be used for cross-platform evaluation, specifically to analize a device's perfomances through QAA or QAOA results. The benchmarking strategy does not rely on previous knowledge of the ideal evolution of the system, but on the quality of the solutions obtained from the QPUs. We have tested the proposed method on \texttt{quera\_aquila} and \texttt{pasqal\_fresnel} with up to 102 qubits, and execution time between $2$ and $4\,\mathrm{\mu s}$. \ar{These benchmarks are scalable because they rely only on the quality of the measured optimization outcomes, rather than on full knowledge of the underlying quantum evolution, making them suitable for application-level performance assessment even when exact quantum simulations become impractical.}

\ar{We emphasize, however, that the proposed protocol is not intended to isolate hardware imperfections from algorithmic limitations. Rather, it provides an end-to-end assessment of application performance under realistic operating conditions. More detailed identification of specific error sources requires complementary techniques such as Hamiltonian learning, device calibration, or analog verification protocols, which address different aspects of processor characterization.}

The relationship between the atomic distances $a$ and the interaction coefficients $C_6$ of the two devices Eq.~\eqref{eq:RelIntDist_Devs} ensures equivalent interaction strengths and thus equivalent Hamiltonians, which enables the cross-platform benchmarking.

Overall, the ability of the QPUs to find valid and good-quality solutions degraded as the problem size increased. For example, for the 11-qubit problem, \texttt{pasqal\_fresnel} gives 63.2\% of valid solutions, while \texttt{quera\_aquila} gives 66.9\%, but for the 85-qubit problem, the valid solutions drop to 4.6\% and 22.1\%, respectively.  
In terms of quality of solutions, \texttt{quera\_aquila} performs better than \texttt{pasqal\_fresnel}, except for the $4\,\mathrm{\mu s}$ QAA schedule at system sizes $N_q = 34, 41$, where \texttt{pasqal\_fresnel} shows slightly better performance. However, the trend does not continue at larger problem sizes \texttt{pasqal\_fresnel} again drops below that of \texttt{quera\_aquila}.
Across all tested conditions, QAOA does not systematically outperform QAA, although for identical evolution times, it performs better on \texttt{quera\_aquila}. Nonetheless, QAOA remains valuable for benchmarking, as it provides insight into how each device handles non-uniform, fast-varying control schedules.

\ar{The benchmark presented here should be viewed as complementary to existing characterization methods for analog quantum processors rather than as a replacement for them. Protocols based on Hamiltonian learning, state tomography, or analog verification provide detailed information about specific physical properties of the hardware, whereas our approach evaluates the processor from an application perspective. In this sense, the benchmark is analogous to end-to-end performance metrics commonly employed in classical computing, where the objective is to quantify the practical ability of a system to solve representative workloads rather than to diagnose individual hardware components.}

\ar{Consequently, a reduced benchmark score may originate from multiple factors, including hardware noise, imperfect calibration, limited algorithmic performance, or a combination thereof. The present protocol intentionally does not attempt to separate these contributions. Instead, it provides a practical and scalable measure of the overall capability of a neutral atom processor to execute native optimization tasks, making it particularly suitable for comparing different hardware generations, control strategies, and algorithmic implementations under identical benchmark conditions.}

Finally, since the QAOA schedule used here was optimized via ideal emulations, an interesting direction for future work would be to explore whether further improvements can be achieved by optimizing the parameters in the presence of noise or by employing variational hybrid loops that combine classical optimization with quantum resources. Similarly, variational methods could also be applied to enhance QAA performance.

\section*{Data Availability}
All problem instances are available at \url{https://jugit.fz-juelich.de/juniq/benchmark-suite/mis-udg}, and results analyzed in this study are available at \url{https://github.com/alejomonbar/Benchmarking-neutral-atom-QPUs}.

\begin{acknowledgments}
\vspace{-10pt}

J. A. Montanez-Barrera and A. B. Rava acknowledge support from the project Jülich UNified Infrastructure for Quantum computing (JUNIQ) that has received funding from the German Federal Ministry of Research, Technology, and Space (BMFTR) and the Ministry of Culture and Science of the State of North Rhine-Westphalia (MKW-NRW). A. B. Rava acknowledges support from the project HPCQS (101018180) of the European High-Performance Computing Joint Undertaking (EuroHPC JU). J. A. Montanez-Barrera acknowledges support from the project EPIQ funded by MKW-NRW.
The authors gratefully acknowledge the Gauss Centre for Supercomputing e.V. (www.gauss-centre.eu) for funding this project by providing computing time on the GCS Supercomputer JUWELS Booster at Jülich Supercomputing Centre (JSC).

We acknowledge support from Amazon Web Services (AWS) through the provision of Amazon Braket credits used for quantum computing experiments on \texttt{quera\_aquila}.
 
\end{acknowledgments}
\clearpage 
% \bibliography{References}
\bibliography{biblio.bib}

@misc{pichler2018quantumoptimizationmaximumindependent,
      title={Quantum Optimization for Maximum Independent Set Using Rydberg Atom Arrays}, 
      author={Hannes Pichler and Sheng-Tao Wang and Leo Zhou and Soonwon Choi and Mikhail D. Lukin},
      year={2018},
      eprint={1808.10816},
      archivePrefix={arXiv},
      primaryClass={quant-ph},
      url={https://arxiv.org/abs/1808.10816}, 
}

@article{Finzgar2024Bayesian,
  title={Designing Quantum Annealing Schedules using Bayesian Optimization},
  author={Fin{\v{z}}gar, Jernej Rudi and Schuetz, Martin J. A. and Brubaker, J. Kyle and Nishimori, Hidetoshi and Katzgraber, Helmut G.},
  journal={Phys. Rev. Research},
  volume={6},
  pages={023063},
  year={2024}
}

@article{Andrist2023MIS,
  title={Hardness of the Maximum Independent Set Problem on Unit-Disk Graphs and Prospects for Quantum Speedups},
  author={Andrist, Ruben S. and Schuetz, Martin J. A. and Minssen, Pierre and others},
  journal={Phys. Rev. Research},
  volume={5},
  pages={043277},
  year={2023}
}

@article{Schuetz2025Compilation,
  title={Quantum Compilation Toolkit for Rydberg Atom Arrays with Implications for Problem Hardness and Quantum Speedups},
  author={Schuetz, Martin J. A. and Andrist, Ruben S. and Salton, Grant and others},
  journal={Phys. Rev. Research},
  volume={7},
  pages={033107},
  year={2025}
}

@article{Ebadi_2022,
   title={Quantum optimization of maximum independent set using Rydberg atom arrays},
   volume={376},
   ISSN={1095-9203},
   url={http://dx.doi.org/10.1126/science.abo6587},
   DOI={10.1126/science.abo6587},
   number={6598},
   journal={Science},
   publisher={American Association for the Advancement of Science (AAAS)},
   author={Ebadi, S. and Keesling, A. and Cain, M. and Wang, T. T. and Levine, H. and Bluvstein, D. and Semeghini, G. and Omran, A. and Liu, J.-G. and Samajdar, R. and Luo, X.-Z. and Nash, B. and Gao, X. and Barak, B. and Farhi, E. and Sachdev, S. and Gemelke, N. and Zhou, L. and Choi, S. and Pichler, H. and Wang, S.-T. and Greiner, M. and Vuletić, V. and Lukin, M. D.},
   year={2022},
   month=jun, pages={1209–1215} }

@misc{hespe2021targetedbranchingmaximumindependent,
      title={Targeted Branching for the Maximum Independent Set Problem}, 
      author={Demian Hespe and Sebastian Lamm and Christian Schorr},
      year={2021},
      eprint={2102.01540},
      archivePrefix={arXiv},
      primaryClass={cs.DS},
      url={https://arxiv.org/abs/2102.01540}, 
}

@article{deraedt87,
	author = {{De Raedt}, Hans},
	doi = {10.1016/0167-7977(87)90002-5},
	journal = {Comp. Phys. Rep.},
	pages = {1},
	title = {{Product formula algorithms for solving the time dependent Schr\"odinger equation}},
	url = {http://www.sciencedirect.com/science/article/pii/0167797787900025},
	volume = {7},
	year = {1987}
}

@article{suzuki1976,
	author = {Suzuki, Masuo},
	day = {01},
	doi = {10.1007/BF01609348},
	issn = {1432-0916},
	journal = {Commun. Math. Phys.},
	month = {Jun},
	number = {2},
	pages = {83--190},
	title = {{Generalized Trotter's formula and systematic approximants of exponential operators and inner derivations with applications to many-body problems}},
	url = {https://doi.org/10.1007/BF01609348},
	volume = {51},
	year = {1976}
}

@article{suzuki84,
	author = {Suzuki, Masuo},
	doi = {10.1063/1.526596},
	journal = {J. Math. Phys.},
	number = {4},
	pages = {601--612},
	title = {{Decomposition formulas of exponential operators and Lie exponentials with some applications to quantum mechanics and statistical physics}},
	url = {https://doi.org/10.1063/1.526596},
	volume = {26},
	year = {1985}
}

@article{trotter59,
	author = {Trotter, H. F.},
	doi = {10.1090/S0002-9939-1959-0108732-6},
	journal = {Proc. Amer. Math. Soc.},
	pages = {545--551},
	publisher = {American Mathematical Society},
	title = {On the product of semi-groups of operators},
	url = {http://www.ams.org/journals/proc/1959-010-04/S0002-9939-1959-0108732-6/home.html#Additional},
	volume = {10},
	year = {1959}
}

@article{huyghebaert1990,
	author = {Huyghebaert, J and {De Raedt}, H },
	doi = {10.1088/0305-4470/23/24/019},
	fjournal = {Journal of Physics A: Mathematical and General},
	journal = {J. Phys. A: Math. Gen.},
	month = {dec},
	number = {24},
	pages = {5777--5793},
	publisher = {{IOP} Publishing},
	title = {{Product formula methods for time-dependent Schr\"odinger problems}},
	url = {https://doi.org/10.1088/0305-4470/23/24/019},
	volume = {23},
	year = {1990}
}

@misc{zunkovič2024VarQAA,
      title={Variational ground-state quantum adiabatic theorem}, 
      author={Bojan Žunkovič and Pietro Torta and Giovanni Pecci and Guglielmo Lami and Mario Collura},
      year={2024},
      eprint={2406.12392},
      archivePrefix={arXiv},
      primaryClass={quant-ph},
      url={https://arxiv.org/abs/2406.12392}, 
}

@article{Saffman2010_QINA,
  title = {Quantum information with Rydberg atoms},
  author = {Saffman, M. and Walker, T. G. and M\o{}lmer, K.},
  journal = {Rev. Mod. Phys.},
  volume = {82},
  issue = {3},
  pages = {2313--2363},
  numpages = {0},
  year = {2010},
  month = {Aug},
  publisher = {American Physical Society},
  doi = {10.1103/RevModPhys.82.2313},
  url = {https://link.aps.org/doi/10.1103/RevModPhys.82.2313}
}

@article{Browaeys_2020MBPNA,
   title={Many-body physics with individually controlled Rydberg atoms},
   volume={16},
   ISSN={1745-2481},
   url={http://dx.doi.org/10.1038/s41567-019-0733-z},
   DOI={10.1038/s41567-019-0733-z},
   number={2},
   journal={Nature Physics},
   publisher={Springer Science and Business Media LLC},
   author={Browaeys, Antoine and Lahaye, Thierry},
   year={2020},
   month=jan, pages={132–142} }

@article{Wagner_2024BenchNA,
   title={Benchmarking a neutral-atom quantum computer},
   volume={22},
   ISSN={1793-6918},
   url={http://dx.doi.org/10.1142/S0219749924500011},
   DOI={10.1142/s0219749924500011},
   number={04},
   journal={International Journal of Quantum Information},
   publisher={World Scientific Pub Co Pte Ltd},
   author={Wagner, N. and Poole, C. and Graham, T. M. and Saffman, M.},
   year={2024},
   month=feb }

@article{Henriet_2020Pasqal,
   title={Quantum computing with neutral atoms},
   volume={4},
   ISSN={2521-327X},
   url={http://dx.doi.org/10.22331/q-2020-09-21-327},
   DOI={10.22331/q-2020-09-21-327},
   journal={Quantum},
   publisher={Verein zur Forderung des Open Access Publizierens in den Quantenwissenschaften},
   author={Henriet, Loïc and Beguin, Lucas and Signoles, Adrien and Lahaye, Thierry and Browaeys, Antoine and Reymond, Georges-Olivier and Jurczak, Christophe},
   year={2020},
   month=sep, pages={327} }

@misc{wurtz2023AquilaQuEra,
      title={Aquila: QuEra's 256-qubit neutral-atom quantum computer}, 
      author={Jonathan Wurtz and Alexei Bylinskii and Boris Braverman and Jesse Amato-Grill and Sergio H. Cantu and Florian Huber and Alexander Lukin and Fangli Liu and Phillip Weinberg and John Long and Sheng-Tao Wang and Nathan Gemelke and Alexander Keesling},
      year={2023},
      eprint={2306.11727},
      archivePrefix={arXiv},
      primaryClass={quant-ph},
      url={https://arxiv.org/abs/2306.11727}, 
}

@misc{blumekohout2013Tomography,
      title={Robust, self-consistent, closed-form tomography of quantum logic gates on a trapped ion qubit}, 
      author={Robin Blume-Kohout and John King Gamble and Erik Nielsen and Jonathan Mizrahi and Jonathan D. Sterk and Peter Maunz},
      year={2013},
      eprint={1310.4492},
      archivePrefix={arXiv},
      primaryClass={quant-ph},
      url={https://arxiv.org/abs/1310.4492}, 
}

@article{Magesan2011BenchQP,
  title = {Scalable and Robust Randomized Benchmarking of Quantum Processes},
  author = {Magesan, Easwar and Gambetta, J. M. and Emerson, Joseph},
  journal = {Phys. Rev. Lett.},
  volume = {106},
  issue = {18},
  pages = {180504},
  numpages = {4},
  year = {2011},
  month = {May},
  publisher = {American Physical Society},
  doi = {10.1103/PhysRevLett.106.180504},
  url = {https://link.aps.org/doi/10.1103/PhysRevLett.106.180504}
}

@article{Ebadi_2021QPM_QUERA,
   title={Quantum phases of matter on a 256-atom programmable quantum simulator},
   volume={595},
   ISSN={1476-4687},
   url={http://dx.doi.org/10.1038/s41586-021-03582-4},
   DOI={10.1038/s41586-021-03582-4},
   number={7866},
   journal={Nature},
   publisher={Springer Science and Business Media LLC},
   author={Ebadi, Sepehr and Wang, Tout T. and Levine, Harry and Keesling, Alexander and Semeghini, Giulia and Omran, Ahmed and Bluvstein, Dolev and Samajdar, Rhine and Pichler, Hannes and Ho, Wen Wei and Choi, Soonwon and Sachdev, Subir and Greiner, Markus and Vuletić, Vladan and Lukin, Mikhail D.},
   year={2021},
   month=jul, pages={227–232} }

@article{Keating_2013,
   title={Adiabatic quantum computation with Rydberg-dressed atoms},
   volume={87},
   ISSN={1094-1622},
   url={http://dx.doi.org/10.1103/PhysRevA.87.052314},
   DOI={10.1103/physreva.87.052314},
   number={5},
   journal={Physical Review A},
   publisher={American Physical Society (APS)},
   author={Keating, Tyler and Goyal, Krittika and Jau, Yuan-Yu and Biedermann, Grant W. and Landahl, Andrew J. and Deutsch, Ivan H.},
   year={2013},
   month=may }

@misc{tibaldi2025analogqaoabayesianoptimisation,
      title={Analog QAOA with Bayesian Optimisation on a neutral atom QPU}, 
      author={Simone Tibaldi and Lucas Leclerc and Davide Vodola and Edoardo Tignone and Elisa Ercolessi},
      year={2025},
      eprint={2501.16229},
      archivePrefix={arXiv},
      primaryClass={quant-ph},
      url={https://arxiv.org/abs/2501.16229}, 
}

@misc{wurtz2023aquila,
    title={Aquila: QuEra's 256-qubit neutral-atom quantum computer},
    author={Jonathan Wurtz and Alexei Bylinskii and Boris Braverman and Jesse Amato-Grill and Sergio H. Cantu and Florian Huber and Alexander Lukin and Fangli Liu and Phillip Weinberg and John Long and Sheng-Tao Wang and Nathan Gemelke and Alexander Keesling},
    year={2023},
    eprint={2306.11727},
    archivePrefix={arXiv},
    primaryClass={quant-ph}
}

@misc{Pasqal2022HPCQS,
  title        = {PASQAL Fresnel},
  url = {https://www.pasqal.com/newsroom/with-two-100-qubits-quantum-computers-from-pasqal-fzj-and-genci-boost-hpcqs-the-pan-european-hybrid-hpc-quantum-infrastructure/},
  note         = {Accessed: 2025-11-05},
  month        = may,
  year         = 2022,
  day          = 30
}

@article{Nelder,
author = {McKinnon, K. I. M.},
title = {Convergence of the Nelder--Mead Simplex Method to a Nonstationary Point},
journal = {SIAM Journal on Optimization},
volume = {9},
number = {1},
pages = {148-158},
year = {1998},
doi = {10.1137/S1052623496303482},
URL = {https://doi.org/10.1137/S1052623496303482},
eprint = {https://doi.org/10.1137/S1052623496303482}
}

@misc{wurtz2023aquilaqueras256qubitneutralatom,
      title={Aquila: QuEra's 256-qubit neutral-atom quantum computer}, 
      author={Jonathan Wurtz and Alexei Bylinskii and Boris Braverman and Jesse Amato-Grill and Sergio H. Cantu and Florian Huber and Alexander Lukin and Fangli Liu and Phillip Weinberg and John Long and Sheng-Tao Wang and Nathan Gemelke and Alexander Keesling},
      year={2023},
      eprint={2306.11727},
      archivePrefix={arXiv},
      primaryClass={quant-ph},
      url={https://arxiv.org/abs/2306.11727}, 
}

@article{Monta_ez_Barrera_2025,
   title={Transfer learning of optimal QAOA parameters in combinatorial optimization},
   volume={24},
   ISSN={1573-1332},
   url={http://dx.doi.org/10.1007/s11128-025-04743-4},
   DOI={10.1007/s11128-025-04743-4},
   number={5},
   journal={Quantum Information Processing},
   publisher={Springer Science and Business Media LLC},
   author={Montañez-Barrera, J. A. and Willsch, Dennis and Michielsen, Kristel},
   year={2025},
   month=may }

@article{Bluvstein_2022QP_Neut,
   title={A quantum processor based on coherent transport of entangled atom arrays},
   volume={604},
   ISSN={1476-4687},
   url={http://dx.doi.org/10.1038/s41586-022-04592-6},
   DOI={10.1038/s41586-022-04592-6},
   number={7906},
   journal={Nature},
   publisher={Springer Science and Business Media LLC},
   author={Bluvstein, Dolev and Levine, Harry and Semeghini, Giulia and Wang, Tout T. and Ebadi, Sepehr and Kalinowski, Marcin and Keesling, Alexander and Maskara, Nishad and Pichler, Hannes and Greiner, Markus and Vuletić, Vladan and Lukin, Mikhail D.},
   year={2022},
   month=apr, pages={451–456} }

@article{browaeys2020many,
  title={Many-body physics with individually controlled Rydberg atoms},
  author={Browaeys, Antoine and Lahaye, Thierry},
  journal={Nature Physics},
  volume={16},
  number={2},
  pages={132--142},
  year={2020},
  publisher={Nature Publishing Group UK London}
}

@article{xu2023constantoverhead,
  title={Constant-overhead fault-tolerant quantum computation with reconfigurable atom arrays},
  author={Qian Xu and J. Pablo Bonilla Ataides and Christopher A. Pattison and Nithin Raveendran and Dolev Bluvstein and Jonathan Wurtz and Bane Vasic and Mikhail D. Lukin and Liang Jiang and Hengyun Zhou},
  journal={Nature Physics},
  volume={20},
  pages={1084--1090},
  year={2024},
  publisher={Nature Publishing Group UK London}
}

@article{Zhou_2025,
   title={Low-overhead transversal fault tolerance for universal quantum computation},
   volume={646},
   ISSN={1476-4687},
   url={http://dx.doi.org/10.1038/s41586-025-09543-5},
   DOI={10.1038/s41586-025-09543-5},
   number={8084},
   journal={Nature},
   publisher={Springer Science and Business Media LLC},
   author={Zhou, Hengyun and Zhao, Chen and Cain, Madelyn and Bluvstein, Dolev and Maskara, Nishad and Duckering, Casey and Hu, Hong-Ye and Wang, Sheng-Tao and Kubica, Aleksander and Lukin, Mikhail D.},
   year={2025},
   month=sep, pages={303–308} }

@article{Bombieri_2025,
   title={Quantum Adiabatic Optimization with Rydberg Arrays: Localization Phenomena and Encoding Strategies},
   volume={6},
   ISSN={2691-3399},
   url={http://dx.doi.org/10.1103/PRXQuantum.6.020306},
   DOI={10.1103/prxquantum.6.020306},
   number={2},
   journal={PRX Quantum},
   publisher={American Physical Society (APS)},
   author={Bombieri, Lisa and Zeng, Zhongda and Tricarico, Roberto and Lin, Rui and Notarnicola, Simone and Cain, Madelyn and Lukin, Mikhail D. and Pichler, Hannes},
   year={2025},
   month=apr }

@misc{schuetz2025qredumisquantuminformedreductionalgorithm,
      title={qReduMIS: A Quantum-Informed Reduction Algorithm for the Maximum Independent Set Problem}, 
      author={Martin J. A. Schuetz and Romina Yalovetzky and Ruben S. Andrist and Grant Salton and Yue Sun and Rudy Raymond and Shouvanik Chakrabarti and Atithi Acharya and Ruslan Shaydulin and Marco Pistoia and Helmut G. Katzgraber},
      year={2025},
      eprint={2503.12551},
      archivePrefix={arXiv},
      primaryClass={quant-ph},
      url={https://arxiv.org/abs/2503.12551}, 
}

@article{Sales_Rodriguez_2025,
   title={Experimental demonstration of logical magic state distillation},
   volume={645},
   ISSN={1476-4687},
   url={http://dx.doi.org/10.1038/s41586-025-09367-3},
   DOI={10.1038/s41586-025-09367-3},
   number={8081},
   journal={Nature},
   publisher={Springer Science and Business Media LLC},
   author={Sales Rodriguez, Pedro and Robinson, John M. and Jepsen, Paul Niklas and He, Zhiyang and Duckering, Casey and Zhao, Chen and Wu, Kai-Hsin and Campo, Joseph and Bagnall, Kevin and Kwon, Minho and Karolyshyn, Thomas and Weinberg, Phillip and Cain, Madelyn and Evered, Simon J. and Geim, Alexandra A. and Kalinowski, Marcin and Li, Sophie H. and Manovitz, Tom and Amato-Grill, Jesse and Basham, James I. and Bernstein, Liane and Braverman, Boris and Bylinskii, Alexei and Choukri, Adam and DeAngelo, Robert J. and Fang, Fang and Fieweger, Connor and Frederick, Paige and Haines, David and Hamdan, Majd and Hammett, Julian and Hsu, Ning and Hu, Ming-Guang and Huber, Florian and Jia, Ningyuan and Kedar, Dhruv and Kornjača, Milan and Liu, Fangli and Long, John and Lopatin, Jonathan and Lopes, Pedro L. S. and Luo, Xiu-Zhe and Macrì, Tommaso and Marković, Ognjen and Martínez-Martínez, Luis A. and Meng, Xianmei and Ostermann, Stefan and Ostroumov, Evgeny and Paquette, David and Qiang, Zexuan and Shofman, Vadim and Singh, Anshuman and Singh, Manuj and Sinha, Nandan and Thoreen, Henry and Wan, Noel and Wang, Yiping and Waxman-Lenz, Daniel and Wong, Tak and Wurtz, Jonathan and Zhdanov, Andrii and Zheng, Laurent and Greiner, Markus and Keesling, Alexander and Gemelke, Nathan and Vuletić, Vladan and Kitagawa, Takuya and Wang, Sheng-Tao and Bluvstein, Dolev and Lukin, Mikhail D. and Lukin, Alexander and Zhou, Hengyun and Cantú, Sergio H.},
   year={2025},
   month=jul, pages={620–625} }

@misc{QuEraComputing,
  title        = {QuEra Computing Inc.},
  howpublished = {\url{https://www.quera.com/}},
  note         = {Accessed: 2025-11-05},
  year         = {2025}
}

@misc{Pasqal,
  title        = {Pasqal - Quantum Processing with Neutral Atoms},
  howpublished = {\url{https://www.pasqal.com/}},
  note         = {Accessed: 2025-11-05},
  year         = {2025}
}

@misc{AtomComputingTech,
  title       = {Atom Computing, Inc.},
  howpublished = {\url{https://atom-computing.com/quantum-computing-technology/}},
  note         = {Accessed: 2025-11-05},
  year         = {2025}
}

@misc{Infleqtion,
  title       = {Infleqtion},
  howpublished = {\url{https://infleqtion.com/}},
  note         = {Accessed: 2025-11-05},
  year         = {2025}
}

@article{Knill_2008,
   title={Randomized benchmarking of quantum gates},
   volume={77},
   ISSN={1094-1622},
   url={http://dx.doi.org/10.1103/PhysRevA.77.012307},
   DOI={10.1103/physreva.77.012307},
   number={1},
   journal={Physical Review A},
   publisher={American Physical Society (APS)},
   author={Knill, E. and Leibfried, D. and Reichle, R. and Britton, J. and Blakestad, R. B. and Jost, J. D. and Langer, C. and Ozeri, R. and Seidelin, S. and Wineland, D. J.},
   year={2008},
   month=jan }

@article{Boixo_2018,
   title={Characterizing quantum supremacy in near-term devices},
   volume={14},
   ISSN={1745-2481},
   url={http://dx.doi.org/10.1038/s41567-018-0124-x},
   DOI={10.1038/s41567-018-0124-x},
   number={6},
   journal={Nature Physics},
   publisher={Springer Science and Business Media LLC},
   author={Boixo, Sergio and Isakov, Sergei V. and Smelyanskiy, Vadim N. and Babbush, Ryan and Ding, Nan and Jiang, Zhang and Bremner, Michael J. and Martinis, John M. and Neven, Hartmut},
   year={2018},
   month=apr, pages={595–600} }

@article{Nielsen_2021,
   title={Gate Set Tomography},
   volume={5},
   ISSN={2521-327X},
   url={http://dx.doi.org/10.22331/q-2021-10-05-557},
   DOI={10.22331/q-2021-10-05-557},
   journal={Quantum},
   publisher={Verein zur Forderung des Open Access Publizierens in den Quantenwissenschaften},
   author={Nielsen, Erik and Gamble, John King and Rudinger, Kenneth and Scholten, Travis and Young, Kevin and Blume-Kohout, Robin},
   year={2021},
   month=oct, pages={557} }

@article{bernien2017probing,
  title={Probing many-body dynamics on a 51-atom quantum simulator},
  author={Bernien, Hannes and Schwartz, Sylvain and Keesling, Alexander and Levine, Harry and Omran, Ahmed and Pichler, Hannes and Choi, Soonwon and Zibrov, Alexander S and Endres, Manuel and Greiner, Markus and others},
  journal={Nature},
  volume={551},
  number={7682},
  pages={579--584},
  year={2017},
  publisher={Nature Publishing Group UK London}
}

@article{bluvstein2021controlling,
  title={Controlling quantum many-body dynamics in driven Rydberg atom arrays},
  author={Bluvstein, Dolev and Omran, Ahmed and Levine, Harry and Keesling, Alexander and Semeghini, Giulia and Ebadi, Sepehr and Wang, Tout T and Michailidis, Alexios A and Maskara, Nishad and Ho, Wen Wei and others},
  journal={Science},
  volume={371},
  number={6536},
  pages={1355--1359},
  year={2021},
  publisher={American Association for the Advancement of Science}
}

@article{cheng2024emergent,
  title={Emergent U (1) lattice gauge theory in Rydberg atom arrays},
  author={Cheng, Yanting and Zhai, Hui},
  journal={Nature Reviews Physics},
  volume={6},
  number={9},
  pages={566--576},
  year={2024},
  publisher={Nature Publishing Group UK London}
}

\clearpage  % Start on a new page
\onecolumngrid  % Switch to one column for the supplementary material

\end{document}